\documentclass[prb,twocolumn,showpacs,preprintnumbers,amsmath,amssymb]{revtex4}
\usepackage{amsfonts}
\usepackage{latexsym}
\usepackage{graphicx}
\usepackage{dsfont}
\usepackage{epsfig}

\newcommand{\Eqref}[1]{Eq.~(\ref{#1})}

\newcommand{\nn}{\nonumber}
\newcommand{\be}{\begin{equation}}
\newcommand{\ee}{\end{equation}}
\begin{document}
\title{Boltzmann equation for fluctuation Cooper pairs in Lawrence-Doniach model.
Possible out-of-plane negative differential conductivity}

\author{Todor~M.~Mishonov}
\email[E-mail: ]{mishonov@phys.uni-sofia.bg}
\author{Yana~G.~Maneva}
\email[E-mail: ]{yanamaneva@gmail.com}
\affiliation{Department of Theoretical Physics, Faculty of Physics,\\
University of Sofia St.~Kliment Ohridski,\\
5 J. Bourchier Boulevard, BG-1164 Sofia, Bulgaria}

\date{\today}

\begin{abstract}
The differential conductivity for the out-of-plane transport in
layered cuprates is calculated for Lawrence-Doniach model in the
framework of time-dependent Ginzburg-Landau (TDGL) theory. The TDGL
equation for the superconducting order parameter is solved in the
presence of Langevin external noise, describing the birth of
fluctuation Cooper pairs. The TDGL correlator of the superconducting
order parameter is calculated in momentum representation and it is
shown that the so defined number of particles obeys the Boltzmann
equation. The fluctuation conductivity is given by an integral over
the Josephson phase $\theta$ of the particles distribution,
depending on that phase, $n(\theta)$ and their velocity $v(\theta)$.
It is demonstrated that in case of overcooling under $T_c$ the
transition from normal to superconducting phase while reducing the
external electric field runs via annulation of the differential
conductivity. The presented results can be used for analysis and
experimental data processing of measurements of differential
conductivity and fluctuation current in strong electric and magnetic
fields. The possible usage of a negative differential conductivity
(NDC) for generation of THz oscillations is shortly discussed.
\end{abstract}

\maketitle
%%%%%%%%%%%%%%%%%%%%%%%%%%%%%%%%%%%%%%%%%%%%%%%%%%%%%%%%%%%%%%%%%%%%%%%%
\section{Introduction}
Fluctuation phenomena in high-temperature cuprates (CuO$_2$) are greatly
pronounced because of the short length of coherence.
The number of degrees of freedom in the sample's
volume is $\mathcal{V}/\xi_{ab}^2(0)\xi_c(0).$ The fluctuations in strongly
anisotropic superconductors are approximately
two-dimensional in layers at a distance $s$ from each other and the number of
fluctuation modes is $\mathcal{V}/\xi_{ab}^2(0)s.$ A reduction of the
dimension, as a rule, leads to enhancement of the fluctuation effects.
The paraconductivity of Aslamazov-Larkin (AL), when to the normal
conductivity we have to add the contribution of the fluctuation
Cooper pairs is the most thoroughly investigated fluctuation phenomenon.
The AL conductivity is easier to be observed if the
normal phase conductivity is small. That is the case of a
strongly disordered conventional superconductor or high-temperature
cuprates, containing CuO$_2$ planes as a main structural detail. The
amplitudes of the electron hopping between CuO$_2$ layers are small,
which leads to intensive diminution of the conductivity in the so
called dielectric $z$-direction, perpendicular to the CuO$_2$ layers.
As a consequence one could apply considerable electric fields
that will not result in significant heating of the sample. The small
heating gives the opportunity to investigate how the strong electric
field's nonlinear effects influence the fluctuation conductivity $\sigma.$ Such
investigations could reveal new details in the kinetics of the
superconducting order parameter. They could specify the parameters of
time-dependent Ginzburg-Landau (TDGL) theory which describes, with
acceptable accuracy, the fluctuation phenomena in
almost all superconductors.
We recommend the monograph by Larkin and Varlamov\cite{Larkin-Varlamov:05}
for a general review on fluctuation phenomena in superconductors and the review
article (and references therein)\cite{Mishonov-Penev:00} which is
especially devoted to Gaussian fluctuations in layered
superconductors. In addition, the recent experimental investigations on
cut-off effects\cite{recent} and reduction of paraconductivity with
increase of the electric field in cuprates\cite{Lang:05} are also
recommended.

The purpose of the current work is to demonstrate that negative
differential conductivity (NDC) can be reached in the specific
nonequilibrium conditions of a supercooled below $T_c$
superconductor in normal state in external electric field $E_c.$
This NDC can find possible applications in the usage of layered
cuprates as active media in THz generators. We derived explicit
formulas for differential conductivity in out of plane direction and
for the current in case of external magnetic field $B_z.$ In the
special case of evanescent magnetic field our results agree with the
work by Puica and Lang.\cite{Puica-Lang:06} Our TDGL result for the
momentum distribution of the fluctuation Cooper pairs obey the
Boltzmann equation. The new results for the current $j_z$ can be
used for the experimental data processing and interpretation of the
experimental results for out-of-plane fluctuation conductivity for
samples in the geometry of plane capacitor with a layered
superconductor between the plates. Special attention is paid to the
analysis of formulas for the differential conductivity in different
physical conditions. How to reach easily the regime of high
frequency oscillations is briefly considered.

For this sake the time dependence of the order parameter is
estimated under Langevin approach and its correlator in coinciding
time arguments is calculated. We have demonstrated that this
correlator in case of homogeneous electric field obeys the master
equation. In this way Boltzmann equation for the fluctuation Cooper
pairs has been derived, which solution gives the momentum
distribution of the superconducting order parameter. The electric
current is represented as an integral over this distribution and the
velocity. The derived formula for the current could be generalized
for a self-consistent treatment of the interaction between the
fluctuations. Such a possibility has been discussed in short. We
have shown that when $T<T_c$ the phase transition to the
superconducting state passes through the critical point of zero
differential conductivity $\sigma_{\mathrm {diff}}=0.$ This
qualitative result is one of the important consequences predicted by
the theory which can be described by the deduced formula for the
current $j_z(\epsilon,\mathcal{P},h)$ as a function of the
dimensionless electric field $\mathcal{P}=eE_z\tau_0s/\hbar$,
dimensionless magnetic field $h=2\pi\xi_{ab}^2(0)B_z/\Phi_0$ and the
reduced temperature $\epsilon\equiv\ln (T/T_c)\approx(T-T_c)/T_c$
Eq.~(\ref{j_final_h}). Further decrease of the electric field $E$ in
the regime $T<T_c$ could originate either electric oscillations or
lead to formation of domain structure, but for sure the homogeneous
phase loses stability and a new physics takes place. What we do is
to offer systematical measurements of the generation of harmonics
for investigation the role of strong electric fields. The
theoretical analysis is completed in the framework of the TDGL
theory applied to the Lawrence-Doniach (LD) model for a layered
superconductor, introduced in the next section.

%%%%%%%%%%%%%%%%%%%%%%%%%%%%%%%%%%%%%%%%%%%%%%%%%%%%%%%%%%%%%%%%%%%%%%%%
\section{Model}

Our starting point is the TDGL equation, which in momentum space
takes the form
\be
\left[\varepsilon (\mathbf{P} - e^{*}\mathbf{A})+a_0\epsilon\right]\psi_p(t)
= -2a_0\tau_0 [\mathrm{d}_t\psi_p(t) - \zeta_p(t)],
\ee
where the argument of the kinetic energy is the
gauge invariant kinetic momentum
\be
\mathbf p(t) = \mathbf P - e^*\mathbf A(t),
\ee
$t$ is the time, and $\tau_0$ is the TDGL relaxation time.
The canonic one $\mathbf P=\mathrm{const}$ is a conserving quantity for
a free particle moving in an external electromagnetic field.
In LD model the energy spectrum is given by
\be \label{E_kin} \varepsilon \left(\mathbf{p}\right)
= \frac{p^2_{x}+p^2_{y}}{2m_{ab}} +
\frac{a_0r}{2}(1-\cos\theta). \ee
Here we have used standard notations for the time-dependent vector
potential of the homogeneous electric field perpendicular to the
layers in our analysis $E_z(t)=-\mathrm{d}_tA_z(t)$, the charge of
the Cooper pairs $|e^{*}|= 2|e|$, their effective mass in the plane
of the layers for the uniaxial superconductors which we study
$m_{ab},$ the effective mass for motion in $c$-direction $m_c$ and
the GL parameter
\be
a_0=\frac{\hbar^2}{2m_{ab}\xi^2_{ab}(0)}=\frac{\hbar^2}{2m_{c}\xi^2_{c}(0)},
\quad
\frac{a_0r}{2}=\frac{\hbar^2}{m_cs^2}.
\ee
For high values of the LD parameter $r={(2\xi_c(0)/s)}^2$ their
model simply describes the anisotropic GL theory, applied to an
uniaxial crystal ($m_a=m_b\ll m_c$). Conversely, when the coherence
length in perpendicular direction $\xi_c(0)$ is much smaller than
the distance between the layers we have a system of two-dimensional
superconductors with a Josephson coupling.

The gauge invariant Josephson's phase
\be \label{Jphase}
\theta(t) = \frac{p_{z}s}{\hbar}=
\frac{P_z - e^{*}A_z(t)}{\hbar/s}
\in (-\pi, +\pi)
\ee
actually is the kinetic momentum of the Cooper pair measured in
units, connected to the lattice constant in $z$ direction $s$. For
the sake of simplicity from now on we are going to omit the $z$
index, which would be understood by the context. For the relaxation
time $\tau_0$ in the TDGL equation, the Bardeen-Cooper-Schrieffer
(BCS) theory gives
\be
\label{time_const}
\frac{\hbar}{2\tau_0}=\frac{8T_c}{\pi}.
\ee
This result for the microscopic theory is extremely robust. The same
value we have for dirty bulk superconductors with isotropic gap and
clean two-dimensional d-type
superconductors.\cite{Larkin-Varlamov:05} Experimental researches in
high-$T_c$ cuprates confirmed this value within the experimental
accuracy. For a discussion of different methods for determination of
the relaxation time $\tau_0,$ and references of relevant
experimental works see Sec. 4.2 and Sec. 4.3 of the review
Ref.~[\onlinecite{Mishonov-Penev:00}]. That is why TDGL theory does
not need modifications when applied for d-wave layered cuprates.

The GL theory is applicable for small by absolute value reduced
temperature $|\epsilon| \ll 1$, whereas we are going to use the
negative values of this parameter for a description of an overcooled
superconductor plugged in an external electric field. For the
Langevin term in the TDGL theory we have a white noise correlator
\be
\langle\zeta^{*}_{p_1}(t_1)\zeta_{p_2}(t_2)\rangle =
\frac{T}{a_0\tau_0}\delta_{p_1,p_2}\delta(t_1 - t_2).
\ee
Analyzing bulk properties of the material we formally assume
periodic boundary conditions for a great distance $L \gg s, $ so
that the quasi-momentum takes the discrete form
\be
p_z=\frac{2\pi\hbar}{L}(\mathrm{integer})
\in
\left(-\frac{\pi\hbar}{s},\; +\frac{\pi\hbar}{s}\right).
\ee
Let us introduce convenient for this task dimensionless variables:
$u=t/\tau$ for the dimensionless time and
\be \omega (\mathbf{k}, \theta) =
\frac{\varepsilon(\mathbf{p})}{a_0} = k_x^2 + k_y^2 +
\frac{r}{2}\left(1-\cos\theta\right) \ee
for the dimensionless kinetic energy, where
\begin{eqnarray}
\nn k_x&=&\frac{p_{x}\xi_a(0)}{\hbar}, \;\;
%\label{k_y}
k_y= \frac{p_{y}\xi_b(0)}{\hbar}, \;\; k_z=
\frac{p_{z}\xi_c(0)}{\hbar},\\
\label{k_x} \theta &=&2k_z/\sqrt{r},
\end{eqnarray}
are the components of the dimensionless wave-vector and the
Josephson's phase. For the dimensionless noise
$\bar\zeta^*_p(u)\equiv \tau_0\zeta_p(t)$ the correlator looks like
\be \label{n_T} \langle \bar\zeta^*_p(u_1)\bar\zeta_p(u_2) \rangle =
n_{_T}\delta(u_1 - u_2),\quad n_{_T} = \frac{T}{a_0}. \ee
For the sake of simplicity let us first consider the
one-dimensional case of hopping of fluctuation current
through a chain of Josephson junctions. Then the TDGL equation
in terms of dimensionless variables reads
\be
\label{TDGL}
\mathrm{d}_u \psi_q(u) = - \nu(u)\psi_q(u) + \bar\zeta_q(u),
\ee
where $q=P_z\xi_c(0)/\hbar$ is the conserving dimensionless
canonical momentum in $z$ direction, $\bar A(u) =
-\,e^*A(u)\xi_c(0)/\hbar$ is the dimensionless potential momentum
corresponding to the dimensionless electric field
\be
f(u) = e^*E(t)\tau_0\xi_c(0)/\hbar = \mathrm{d}_u \bar A(u).
 \ee
The dimensionless decay rate is given by
\be
\label{velnarazpad}
2\nu(u)\equiv \omega(q+\bar A(u)) + \epsilon
\ee
and the kinetic dimensionless momentum for the one-dimensional task
we are solving is
\be k_z = q + \bar A(u). \ee
%

%%%%%%%%%%%%%%%%%%%%%%%%%%%%%%%%%%%%%%%%%%%%%%%%%%%%%%%%%%%%%%%%%%%%%%%%
\section{Derivation and solution of the Boltzmann equation}

The TDGL equation Eq.~(\ref{TDGL}) is a linear ordinary differential
equation and one can easily check that its solution has the form
\begin{eqnarray}
\nn \psi_q(u)& =& \left\{ \int_{u_2=0}^u \bar\zeta_q(u_2) \right.
\exp\left[ \int_{u_1=0}^{u_2}\nu(u_1)\mathrm{d}u_1\right]u_2 \\
 & & \left. +  \; \psi_q(0)\right\}\exp
\left[ - \int_{u_3=0}^u\nu(u_3)\mathrm d u_3\right].
\end{eqnarray}
The averaging of the superconductor's order parameter over the noise
gives the distribution of the mean number of particles with respect
to the canonical momentum
\begin{eqnarray}
\label{Boleqn}
\nn n_q(u)& =&  \langle \psi^*_q(u)\psi_q(u) \rangle\\
\nn & =& n_{_T} \int^u_{0}\mathrm du_2\exp
\left[- 2\int^u_{u_2}\mathrm d u_1\, \nu(u_1)\right]\\
& &+\; n_q(u=0)\exp\left[-2\int^u_{0}\nu(u_3)\,\mathrm du_3\right]\!\!.
\end{eqnarray}
In case of constant electric field
the vector-potential and the decay rate take the form
\be \label{decay_rate} \bar A(u) = fu,\quad 2\nu(u)=\frac{r}{2}
\left[1-\cos(\phi+2\mathcal Pu)\right] +\epsilon, \ee
where we have introduced the notations
\be \mathcal{P}\equiv \frac{f}{\sqrt r} = \frac{eE_z\tau_0s}{\hbar},
\quad
\phi\equiv\frac{P_zs}{\hbar}=\frac{2}{\sqrt{r}}\,q=\mathrm{const}.
\ee
 An elementary integration of the upper
equation~(\ref{Boleqn}) in the limit $u \rightarrow \infty,$ using
the trigonometric relation
$$\sin(\phi+2\mathcal P\tilde{u})-\sin\phi
= 2\cos\left(\phi+\mathcal P\tilde u\right)\sin\left(\mathcal
P\tilde u\right)$$ gives
\begin{eqnarray}
\label{BolSol} \!\!\!\!\!\!\!\!&&\frac{n(\theta)}{n_{_T}}
=\int_0^\infty d\tilde u\nn\\
\!\!\!\!\!\!\!\!&&\times\exp\left[ -\!
\left(\epsilon + \frac{r}{2}\right)\tilde{u} +
\frac{r}{2\mathcal{P}} \sin(\mathcal{P} \tilde u)
\cos\left(\theta-\mathcal{P}\tilde u\right) \right]\!.
\end{eqnarray}

Here we have turned to a new integration variable $\tilde u =
u-u_2,$ see Eq.~(\ref{Boleqn}), which describes the birth of
fluctuation Cooper pairs in the moment $\tau_0\tilde u$ before the
time of observation $t=\tau_0 u.$ In such a case the integrant gives the
age distribution of fluctuation Cooper pairs. The periodical
dependence $\sin(\theta + 2\mathcal{P}\tilde{u})$ on time $\tilde u$
is a reflection of the Bloch oscillations of the free charged
particles which move in a periodical potential and constant electric
field.

In case of weak electric fields $\sin(\mathcal P\tilde u)\approx
\mathcal{P}\tilde u$ we obtain the equilibrium distribution density
function
\be
\bar n(\theta) =
\frac{n_{_T}}{\frac{r}{2}\left(1-\cos\theta\right)+ \epsilon}.
\ee
The reconstruction of the plane components of the momentum leads to
the general formula of Rayleigh-Jeans for the equilibrium
distribution of the particles
\be
\bar n_p = \frac{T}{\frac{p_{x}^2+p_{y}^2}{2m_{ab}}+
\frac{\hbar^2}{m_c^2s^2}\left(1-\cos\frac{p_{z}s}{\hbar} \right) + a_0\epsilon}
=\frac{T}{\varepsilon(\mathbf p) - \mu}.
\ee
Hereby formally we may identify the chemical potential with the Landau
parameter $\mu = - a_0\epsilon.$ In this sense the critical temperature
$T_c$ sets the point of annulation of the chemical potential and fixes
the beginning of Bose-Einstein condensation of the fluctuation
Cooper pairs.
It may be easily checked that the momentum distribution of the particles
obeys the master equation of Boltzmann
\be
\label{Boltzmann_2006}
\mathrm d_t n_\mathbf P (t) =
-\frac{n_\mathbf P (t) -
\bar n_{\mathbf P - e^*\mathbf A}(t)}{\tau_{\mathbf P - e^*\mathbf A}}
= -\frac{n_\mathbf P (t)}{\tau_{\mathbf P - e^*\mathbf A}}
+\frac{n_{_T}}{\tau_0},
\ee
where the decay rate is proportional to the kinetic energy
measured from the chemical potential
\be
\frac{1}{\tau_{\mathbf P - e^*\mathbf A}}
= \frac{2}{\tau_0}\nu\left(\mathbf P - e^*\mathbf A(t)\right)
= \frac{\varepsilon (\mathbf p)+a_0\epsilon}{a_0\tau_0}.
\ee
In such a way we derived the Boltzmann equation for fluctuation
Coper pairs for out-of-plane transport of a layered superconductor.
The argument of  the distribution and decay rate is the
time-dependent kinetic momentum. The direct derivation of Boltzmann
equation directly from the TDGL theory is too long and contains a
lot of technical details.\cite{Damianov:96}

In case of zero electric field the relaxation rate of the order
parameter with zero value of the wave vector runs through a typical
slowing down
\be
\tau(\epsilon)=
\tau_{p\,=\,0} = \frac{\tau_0}{\epsilon}.
\ee
This is the characteristic time of ``drying'' at $\epsilon>0$ of the
spatially homogeneous Bose condensate with an order parameter
$\psi(\mathbf r)= \mathrm{const}.$ If we consider the distribution
depending on the kinetic momentum we obtain from TDGL theory the
standard form of the Boltzmann equation derived in 1872
\begin{eqnarray}
\label{Boltzmann_1872}
&&\partial_t n(\mathbf p, t) +
e^*\mathbf E(t)\!\cdot\!\partial_{\mathbf p} n(\mathbf p, t)\nn\\
&&= - \frac{n(\mathbf p, t)-\bar n(\mathbf p)}
{\tau(\mathbf p)}
= - \frac{n(\mathbf p, t)         }
{\tau(\mathbf p)}
+ \frac{n_{_T}}{\tau_0}.
\end{eqnarray}
In our dimensionless variables for the static case this equation
reads
\be
2\mathcal P\mathrm d_\theta n(\theta) = -\left[\frac{r}{2}
(1-\cos\theta)+\epsilon\right]n(\theta) + n_{_T},
\ee
which solution is \Eqref{BolSol}.

In such a way we derived the Cooper pair Stoss-integral in the
framework of the TDGL theory, the result is formally an energy
dependent $\tau$-approximation with a constant birth rate in
momentum space
\be
\frac{\bar n(\mathbf p)}
{\tau(\mathbf p)}
=
\frac{n_{_T}}{\tau_0}.
\ee
This integral describes collisions between normal particles and
creation of Cooper pairs and back process of the decay of the
condensate particles. We wish to point out that the TDGL equation is
a diffusion-like equation which does not lead to quasi-classical
dynamics of quasi-particles.\cite{MPI:02} This shows that Boltzmann
equation has broader applicability than arguments used in his
textbook presentations.

Let us now turn back to the analysis of the distribution function.
In the continual limit when the coherence length significantly
exceeds the lattice constant $\xi_c(0) \gg s$ or in other words when
the LD parameter is big enough $r \gg 1$ we have to consider small
Josephson's phase $|\theta| \ll 1$ and weak fields
$|\mathcal{P}\tilde{u}| \ll 1.$ The distribution over the kinetic
momentum Eq.~(\ref{BolSol}) then becomes\cite{MPI:02,MPGAD:03}
\be
n(k_z)=n_{_T} \int_0^{\infty}\exp\left\{-(\epsilon+k_z^2)\tilde u
+ k_zf\tilde u^2 - \frac{1}{3}f^2\tilde u^3\right\}\mathrm d \tilde u.
\ee
While in Refs.~\onlinecite{MPI:02, MPGAD:03} are analyzed bulk
superconductors in the present work we consider a layered
superconductor with Josephson coupling between the layers for which
we apply LD model. The last formula may be derived directly from the
Boltzmann equation applied in case of parabolic dispersion
$\varepsilon (p_{z}) \approx p^2_{z}/2m_c.$ After we have
investigated the momentum distribution here in the next section we
will obtain a general formula for the density of the electrical
current.

%%%%%%%%%%%%%%%%%%%%%%%%%%%%%%%%%%%%%%%%%%%%%%%%%%%%%%%%%%%%%%%%%%%
\section{Fluctuation current}

Let us start our analysis with the one-dimensional case as the results
may be easily extrapolated for the case of higher dimensions.
The density of the current
\be
\label{sum_j}
j_z = \sum_{p_{z}} e^* v(p_{z})\frac{n(p_{z})}{L}
\ee
is a product of the particles' charge $e^*,$ their density
$n(p_{z})/L$ and velocity
\be
v(p_{z}) = \frac{\partial \varepsilon}{\partial p_{z}}
= \frac{\hbar}{m_cs}\sin\theta.
\ee
This formula is a special case of the general procedure for
derivation of the current density in the framework of the GL's
theory (Ref.~[\onlinecite{LLIII}], sec.~115;
Ref.~[\onlinecite{LLIX}], sec.~45)
\be
j(\mathbf r) = -\frac{\delta}{\delta A(r)}\int \psi^*(\mathbf r)\
\varepsilon (-i\hbar\nabla - e^* A(\mathbf r))\psi(\mathbf r)\mathrm d^3 x,
\ee
where after the functional integration the vector potential is
put to be spatially homogeneous and the order parameter takes the
form of a plane wave
\be
\psi(\mathbf r)
= \sqrt{\frac{n_p}{\mathcal V}}e^{i\mathbf p
\cdot \mathbf r/\hbar},\quad n_p
= \int {|\psi(\mathbf r)|}^2\mathrm d^3 x,\quad \mathcal V
= \int \mathrm d^3 x.
\ee
In the one-dimensional case,
imagine current along a chain of Josephson junctions,
if we use the relation
\be
\label{p_z_theta}
\sum_{p_z} = \frac{L}{2\pi\hbar}\int \mathrm dp_z
=L\int_{-\pi}^\pi \frac{\mathrm d \theta}{2\pi s},
\ee
the substitution of the Boltzmann equation's solution
Eq.~(\ref{BolSol}) in the general formula for the current
Eq.~(\ref{sum_j}) gives
\begin{eqnarray}
\label{j_1D_initial}
\nn j^\mathrm{(1D)}_z &=& \frac{e^*rT}{4\pi\hbar}\int^{\pi}_{-\pi}
\sin\theta\mathrm d\theta\int^\infty_0
\exp\left\{-\left(\epsilon + \frac{r}{2}\right)\tilde u \right.\\
&+& \left. \frac{r}{2\mathcal{P}}
\sin(\mathcal{P}u)\cos\left(\theta-\mathcal{P}\tilde u\right)
\right\}\mathrm d\tilde u.
\end{eqnarray}
Here we have taken into account that
$\theta = p_{z}s/\hbar$ and $a_0n_{_T} = T,$
see Eq.~(\ref{Jphase}) and Eq.~(\ref{n_T}).
The averaging with respect to the Josephson's phase $\theta$
can be expressed by the modified Bessel functions\cite{Korn-Korn:61}
\be
I_m(z)=(-1)^m\int_{-\pi}^\pi \frac{\mathrm d \theta}{2\pi}
\cos m\theta \,\mathrm e^{-z\cos\theta}
=i^{-m}J_m(iz).
\ee
Thus the formula for the current reads
\be \label{j_1D} j^\mathrm{(1D)}_z =
\frac{e^*rT}{2\hbar}\int^\infty_0 \mathrm d \tilde u \mathrm
e^{-(\epsilon+r/2)\tilde u}\sin(\mathcal{P}\tilde u)
I_1\!\left(\frac{r\sin(\mathcal{P}\tilde u)}{2\mathcal{P}}\right).
\ee

In order to calculate the out-of-plane component of the current
$j_z$ in the LD model we have to consider the complementary decay
rate excited by the kinetic energy in the $xy$ plane, i.e $ab$
CuO$_2$ plane. For this reason formula Eq.~(\ref{velnarazpad}) has
to be modified as follows
\be
2\nu = \frac{\varepsilon(\mathbf p)+ a_0(\epsilon)}{a_0}
= \frac{r}{2}(1-\cos\theta) + \eta + \epsilon,
\ee
where according to Eq.~(\ref{k_x})
\be
\eta \equiv \frac{p_{x}^2 + p_{y}^2}{2m_{ab}a_0} = k_x^2+k_y^2
\ee
is the dimensionless in-plane kinetic energy limited by a cut-off
parameter $c;$ for more details see, for example,
Ref.~\onlinecite{Mishonov-Penev:00}. We have to sum over the kinetic
momentums in the $ab$-plane
\be
\int\!\!\int\frac{\mathrm dp_{x}\mathrm dp_{y}}{{(2\pi\hbar)}^2}
f\!\left(\frac{p_{x}^2+p_y^2}{2m_{ab}a_0}\right)
\!= \frac{1}{4\pi\xi^2_{ab}(0)}\int^c_0 \mathrm d\eta f(\eta).
\ee
This additional integration with respect to the two-dimensional
degrees of freedom taken into account in Eq.~(\ref{j_1D})
gives the replacement
\be \label{replacement} \mathrm e^{-\epsilon\tilde u} \rightarrow
\frac{\mathrm e^{-\epsilon\tilde u}}{4\pi\xi^2_{ab}(0)} \int_0^c
\mathrm e^{-\eta\tilde u}\mathrm d\eta =\frac{\mathrm
e^{-\epsilon\tilde u}(1-\mathrm e^{c\tilde
u})}{4\pi\xi^2_{ab}(0)\tilde u}. \ee
Then the expression for the fluctuation current
takes the final form
\begin{eqnarray}
\label{j_final}
\nn
j_z(\bar\epsilon,\mathcal{P})
&=& \frac{erT}{4\pi\hbar\xi_{ab}^2(0)}\int_{\tilde u=0}^\infty
\mathrm d\tilde u \mathrm e^{-(\bar\epsilon + r/2)\tilde u}
\frac{1-\mathrm e^{-c\tilde u}}{\tilde u}\\
&& \times \sin(\mathcal{P}\tilde u)
I_1\left(\frac{r\sin(\mathcal{P}\tilde u)}{2\mathcal{P}}\right)\!,
\end{eqnarray}
where $\bar\epsilon$ means the self-consistent treatment of
$\epsilon$ parameter, which will be analyzed in the next section.

%%%%%%%%%%%%%%%%%%%%%%%%%%%%%%%%%%%%%%%%%%%%%%%%%%%%%%%%%%%%%
\section{Maxwell-Hartree self-consistent approximation}
Up to this moment we have neglected the influence of
the nonlinear term in the GL theory. The coefficient
standing in front of this term participates,
for instance, in the equation for the equilibrium value
of the order parameter above $T_c$
\be
[a_0\epsilon + b{|\psi|}^2]\psi = 0.
\ee
The idea of the self-consistent approximation (SCA) is to replace
the density of particles with its value averaged over all
fluctuation modes. In such a way we obtain a self-consistent
equation for the renormalized reduced temperature, see
Ref.~\onlinecite{Mishonov-Penev:00}, Sec.~3.3
\be
\label{chem}
\bar\epsilon = \ln\frac{T}{T_c}
+ \frac{b}{a_0}n^{(3D)}(\bar\epsilon).
\ee
The idea is coming from the Maxwell treatment of the ring of Saturn
-- the first work on collective phenomena in physics. Maxwell
concluded in 1856 that the ring cannot be a rigid object, but
consists of ``indefinite number of unconnected particles'' and each
of them is moving in the averaged gravitational field of the others.
Analogously the fluctuation density of Cooper pairs renormalizes its
chemical potential as shown in Eq.~(\ref{chem}).

At first from Eq.~(\ref{BolSol}) we calculate the one dimensional
density of the Cooper pairs
\be
n^\mathrm{(1D)}=\int_{-\pi}^{\pi}\frac{\mathrm d\theta}{2\pi s}n(\theta).
\ee
Analogously to the calculation of the current
Eqs.~(\ref{sum_j}) and (\ref{p_z_theta})
now we have
\be \label{n_1D} n^\mathrm{(\!1D)}(\bar\epsilon,\mathcal{P})
=\frac{n_T}{s}\int_{0}^{\infty}\mathrm d \tilde u \mathrm
e^{-(\bar\epsilon+r/2)\tilde u}
I_0\left(\frac{r\sin(\mathcal{P}\tilde u)}{2\mathcal{P}}\right). \ee

The volume density can be obtained via summation of this
one-dimensional density with respect to the in-plane degrees of freedom
\be
\label{n_3D}
n^\mathrm{(3D)}(\bar\epsilon) = \int\int
\frac{\mathrm{d}p_x\mathrm{d}p_y}{{(2\pi\hbar)}^2}n^\mathrm{(\!1D)}
\left(\bar\epsilon+
\frac{p^2_{x}+ p^2_{y}}{2m^*_{ab}a_0}
\right)
,
\ee
cf. the analogous transition between 2D and 3D case for in-plane
conductivity of layered superconductors,
Ref.~\onlinecite{Mishonov-Penev:00}, Sec.~2.3. The kinetic energy
cut-off, Ref.~\onlinecite{Mishonov-Penev:00}, Eq.~(11)
\be
\frac{p^2_{x}+ p^2_{y}}{2m^*_{ab}}< a_0c
\ee
leads to the same replacement as in Eq.~(\ref{replacement}) and for the
3D density we get
\begin{eqnarray}
\nn n^{(3D)}(\bar\epsilon,\mathcal{P}) = \frac{n_T}{4\pi
s\xi^2_{ab}(0)}\int_{\tilde u=0}^\infty \mathrm{d} \tilde u
\frac{1-\mathrm e^{-c\tilde u}}{\tilde u}
\\
\times \mathrm e^{-(\bar \epsilon + r/2)\tilde u}
I_0\left(\frac{r\sin\mathcal{P}}{2\mathcal{P}}\right) .
\end{eqnarray}

Additionally we may represent the nonlinear parameter
$b$ in terms of the Ginzburg-Landau parameter
$\kappa_{\mathrm{GL}}=\lambda_{ab}(0)/\xi_{ab}(0)$ and the
flux quantum $\Phi_0,$ see Ref.~\onlinecite{Mishonov-Penev:00}, Eq.~(165)
\be
b= 2\mu_0{\left(\frac{\pi\hbar\kappa_{GL}}{\Phi_0m^*_{ab}}\right)}^2.
\ee
We also introduce the dimensionless Ginzburg number for a layered
superconductor (cf. Ref.~\onlinecite{Mishonov-Penev:00}, Eq.~(168))
\be
\epsilon_\mathrm{Gi}(T)=2\pi\mu_0\frac{T}{s}
\left(\frac{\lambda_{ab}(0)}{\Phi_0}\right)^2
=\frac{2\mu_0\kappa_\mathrm{GL}^2e^2\xi_{ab}^2(0)T}{\pi\hbar^2 s}.
\ee

In terms of the so introduced notations the self-consistent equation
for the renormalized temperature reads
\be \label{Maxwell-Hartree} \bar\epsilon = \epsilon +
\epsilon_{\mathrm{Gi}} \int_0^\infty\mathrm d \tilde u
\frac{1-\mathrm e^{-c\tilde u}}{\tilde u}\mathrm
e^{-(\bar\epsilon+r/2)\tilde u}
I_0\left(\frac{r\sin(\mathcal{P}\tilde u)}{2\mathcal{P}}\right)\!,
\ee
in agreement with the results of Puica and Lang
Ref.\onlinecite{Puica-Lang:06}, Eqs.~(11) and (17), where the
temperature-dependent Ginzburg number $\epsilon_{\mathrm{Gi}}(T)$ is
denoted by $gT.$ The solution $\bar\epsilon$ of equation
(\ref{Maxwell-Hartree}) has to be substituted as an argument in the
formula for the current Eq.~(\ref{j_final}). This agreement
demonstrates that it is possible to derive in the beginning an
1-dimensional (1D) formula for the Josephson chain and later on to
generalize this approach to a 3-dimensional (3D) case of layered
superconductor.
In the next section we will analyze the modification
of this result in case of out-of-plane external magnetic field.

%%%%%%%%%%%%%%%%%%%%%%%%%%%%%%%%%%%%%%%%%%%%%%%%%%%%%%%%%%%%%%%%%%%%%%%%%%
\section{Influence of a perpendicular magnetic field. Magnetoconductivity}

%%%%%%%%%%%%%%%%%%%%%%%%%
\subsection{General case}
Let us consider an external magnetic field also applied
perpendicularly to the CuO$_2$ planes. In this case the variables
can be separated and the task for the calculation of the current
reduces to the 1D case Eq.~(\ref{j_1D}), which we have already
considered. For the two-dimensional movement the magnetic field
arouses equidistant oscillator spectrum
\be
\frac{\textbf{p}_{ab}}{2m_{ab}^*}
\rightarrow \hbar\omega_c\left(n+\frac{1}{2}\right),
\qquad
\omega_c=\frac{e^*B_z}{m^*_{ab}}.
\ee
The degeneration rate is determined by the Landau sub-zone capacity
$B/\Phi_0,$ where $\Phi_0=2\pi\hbar/e^*$ is the magnetic flux quantum.
For parametrization of the GL theory it is convenient to introduce the
linear extrapolated upper critical field, represented via the coherence
length in the $ab$-plane
\be
\label{uppercritfield}
B_{c2}(0)\equiv-T_c\left.\frac{B_{c2}(T)}{d T}\right|_{T_c}
=\frac{\Phi_0}{2\pi\xi^2_{ab}(0)}.
\ee
For facilitation we will work with the dimensionless
magnetic field
\be
h\equiv\frac{B_z}{B_{c2}(0)}=\frac{\hbar\omega_c}{2a_0},
\ee
with the help of which the dimensionless spectrum quantizes as
\be \eta\rightarrow (2n+1)h, \qquad n=0,\;1,\;2,\;3,\;\dots \ee
The maximal kinetic energy is reached at some big number $N_c,$
which cuts off the summation over the Landau levels
\be
\hbar\omega_cN_c=a_0c,
\qquad
2hN_c=c.
\ee
The influence of the magnetic field reduces to the replacement
of the integration over the kinetic energy to a restricted summation
over the Landau levels
\be \int_0^c\mathrm d\eta f(\eta) \rightarrow \Delta\eta
\sum_{n=0}^{N_c-1}f\left((2n+1)h\right). \ee
In particular, in the exponential dependence, which we have in
Eq.~(\ref{replacement}), we have to sum up the limited
geometric progression
\be \int_0^c \mathrm d\eta\, \mathrm e^{-\eta \tilde u} \rightarrow
2h \sum_{n=0}^{N_c-1} \mathrm e^{-(2n+1)h\tilde u}, \ee
which more easily can be represented as a subtraction of two infinite ones
\be
\sum_{n=0}^{N_c-1}=\sum_{n=0}^{\infty}-\sum_{n=N_c}^{\infty}.
\ee
In this way the summation over the in-plane degrees of freedom
simply reduces to the shift
\be \int_0^c \mathrm d\eta\, \mathrm e^{-\eta \tilde u} \rightarrow
\frac{h}{\sinh h\tilde u}\left(1-\mathrm e^{-c\tilde u}\right). \ee
Then the substitution described in Eq.~(\ref{replacement}) in the
presence of external magnetic field takes the form
\be
e^{-\epsilon\tilde u} \rightarrow
\frac{e^{-\epsilon\tilde u}}{4\pi\xi^2_{ab}(0)}
\int_0^c e^{-\eta\tilde u}\mathrm d\eta
\rightarrow
\frac{\mathrm e^{-\epsilon\tilde u}(1-\mathrm e^{c\tilde u})h}
{4\pi\xi^2_{ab}(0)\sinh h\tilde u}.
\ee
This is the recipe with which one may switch from the already
solved one-dimensional task Eq.~(\ref{j_1D}) to the three-dimensional.
Comparison with the case of zero magnetic field shows that we may use
the 3D formulas Eqs.~(\ref{Maxwell-Hartree}) and (\ref{j_final}), where we
only have to make the substitution
\be
\frac{1}{\tilde u}\rightarrow \frac{h}{\sinh (h\tilde u)}.
\ee
In this way, in agreement with Ref.~\onlinecite{Puica-Lang:06}, Eqs.
(13) and (14), we derive the final formulas for the self-consistent
reduced temperature
\be \label{Maxwell-Hartree_h} \bar\epsilon = \epsilon +
\epsilon_{\mathrm{Gi}} \int_0^\infty\!\mathrm d \tilde u \,
\frac{(1-\mathrm e^{-c\tilde u})h}{\sinh h\tilde u}\, \mathrm
e^{-(\bar\epsilon+r/2)\tilde u}
I_0\left(\frac{r\sin(\mathcal{P}\tilde u)}{2\mathcal{P}}\right)\!
\ee
and the perpendicular to the CuO$_2$ planes density of the current
\begin{eqnarray}
\label{j_final_h} \nn j_z(\bar\epsilon,\mathcal{P},h)\! &=&\!
\frac{e^2\xi_{c}^2(0)E_z}{16\hbar s\xi_{ab}^2(0)} \int_{\tilde
u=0}^\infty \mathrm d\tilde u
\frac{(1-\mathrm e^{-c\tilde u})h}
{\sinh h\tilde u}\mathrm e^{-(\bar\epsilon + r/2)\tilde u}\\
&& \times \frac{\sin(\mathcal{P}\tilde u)}{\mathcal{P}}
I_1\left(\frac{r\sin(\mathcal{P}\tilde u)}{2\mathcal{P}}\right)\!,
\end{eqnarray}
where we have used the relation
\be \frac{erT\mathcal P}{4\pi\hbar\xi_{ab}(0)} =
\frac{e^2\xi_{c}^2(0)E_z}{16\hbar s\xi_{ab}^2(0)}. \ee
This explicit formula is one of the new results of the present work.
Let us now analyze the magnetoconductivity in some characteristic
particular examples.

%%%%%%%%%%%%%%%%%%%%%%%%%%%%%%%%%
\subsection{Weak magnetic fields}

Let us consider the application of the upper formulas
in case of weak magnetic field $h\ll|\bar\epsilon|.$
If we use the well-known summation of
Euler-MacLaurin (cf. with Ref.~\onlinecite{Mishonov-Penev:00}, Eq.~(23))
we have
\be
\frac{x}{\sinh x}= 1 -\frac{1}{6}x^2+\frac{7}{360}x^4- \frac{31}{15120}x^6+\dots
\ee
For weak magnetic fields we may use the approximation
\be
\label{razlojzaweakfields}
\frac{h}{\sinh h\tilde u}\approx\frac{1}{\tilde u}-\frac{1}{6}h^2\tilde u.
\ee
The substitution of this decay into Eq.~(\ref{j_final_h}) leads to the special
expression for the current in the presence of weak magnetic fields
\begin{eqnarray}
\label{j_final_h_small}
\nn
j_z(\bar\epsilon,\mathcal{P},h)\! &=&j_z(\bar\epsilon,\mathcal{P})\\
\nn &&\! -\frac{erTh^2}{24\pi\hbar\xi_{ab}^2(0)}\int_{\tilde u=0}^\infty
\mathrm d\tilde u \mathrm e^{-(\bar\epsilon + r/2)\tilde u}
(1-\mathrm e^{-c\tilde u})\tilde u\\
&& \times
\sin(\mathcal{P}\tilde u)
I_1\left(\frac{r\sin(\mathcal{P}\tilde u)}{2\mathcal{P}}\right)\!.
\end{eqnarray}
The coefficient in front of the integral over the reducing term
may be expressed via the coherent lengths and the lattice constant
as follows
\be
\frac{erTh^2}{24\pi\hbar\xi_{ab}^2(0)}
=\frac{2\pi eT\xi^2_c(0)}{3\hbar s^2}
\left(\frac{\xi_{ab}^2(0)B_z}{\Phi_0}\right)^2.
\ee
The experimental data fitting with this formula could lead to more
precise specification of $\xi_c(0)$ for the investigated sample. The
procedure Eq.~(\ref{razlojzaweakfields}) applied to
Eq.~(\ref{Maxwell-Hartree_h}) ends in a self-consistent equation for
the reduced temperature in case of weak magnetic fields
\begin{eqnarray}
\label{Maxwell-Hartree_h_small} \nn \bar\epsilon(h) \approx \epsilon
- \epsilon_{\mathrm{Gi}}\frac{h^2}{6} \int_0^\infty\!&\mathrm d
\tilde u &\, \mathrm e^{-(\bar\epsilon+r/2)\tilde u}
(1-\mathrm e^{-c\tilde u})\\
&\times& \tilde u I_0\!\left(\frac{r\sin\mathcal{P}\tilde
u}{2\mathcal{P}}\right)\!.
\end{eqnarray}
In the next subsection we will analyze the case of values of the
magnetic field close to the phase curve $B_{c2}(T),$ where the
influence of the fluctuation pairs is most vividly expressed.

%%%%%%%%%%%%%%%%%%%%%%%%%%%%%%%%%%%%%%%%%%%%%%%%
\subsection{Strong fields close to $B_{c2}(T)$}

Close to the phase curve it is more convenient to account the reduced
temperature in regards of the phase transition temperature, which is a
function of the external magnetic field $T_{c2}(B).$ Comparison with
Eq.~(\ref{uppercritfield}) reads
\be
\epsilon_h=\frac{T-T_{c2}(B)}{T_c}=\epsilon+h,
\qquad |\epsilon_h|\ll h.
\ee
In the case of strong magnetic fields it is more appropriate to
use the 1D formula Eq.~(\ref{j_1D}), where we have to add a
summation over the Landau levels, taking into account the degeneration
and the cut off
\be j^\mathrm{(3D)}_z(\bar\epsilon,\mathcal{P},h)
\approx\frac{B}{\Phi_0}
\sum^\infty_{n=0}j^\mathrm{(1D)}_z(\bar\epsilon_h+2nh,\mathcal{P}).
\ee
In close vicinity of the upper critical field $|\bar\epsilon|\ll h$ in the
summation the share of the lowest Landau level dominates over the others
and its contribution gives
\be \label{j_3D_B} j^{(3D)}_z
\!\approx\!\frac{e^*rTB}{2\hbar\Phi_0}\!\int^\infty_0
\!\!\!\mathrm d \tilde
u \mathrm e^{-(\bar\epsilon_h+r/2)\tilde u}\sin(\mathcal{P}\tilde u)
I_1\!\left(\!\frac{r\sin(\mathcal{P}\tilde u)}{2\mathcal{P}}\!\right).
\ee
Similar procedure applied to the number of particles' density
Eq.~(\ref{n_1D}), see also Eq.~(\ref{Maxwell-Hartree}),
\be
\bar\epsilon_h=\epsilon_h+\frac{bB}
{a_0\Phi_0}n^\mathrm{(\!1D)}(\bar\epsilon_h,\mathcal{P})
\ee
determines the equation for the self-consistent reduced temperature
\be \label{n_3D} \bar\epsilon_h=\epsilon_h+2h\epsilon_\mathrm{Gi}
\int_{0}^{\infty}\mathrm d \tilde u \mathrm
e^{-(\bar\epsilon_h+r/2)\tilde u}
I_0\left(\frac{r\sin(\mathcal{P}\tilde u)}{2\mathcal{P}}\right). \ee
Close to the upper critical magnetic field the fluctuations are practically
one-dimensional and such a reduction of the space dimensionality significantly
amplifies the fluctuation phenomena.

Next we are going to derive an explicit formula for the differential
conductivity in the most general case of LD model and analyze it in
the particular cases of both weak and strong magnetic fields. The
physical conditions of possible occurrence of negative differential
conductivity will be discussed.

%%%%%%%%%%%%%%%%%%%%%%%%%%%%%%%%%%%%%%%%%%%%%%%%%%%%%%%%%%%%%%%%%%%%%%%
\section{Differential conductivity. Possible NDC}

The non-linear effects over the conductivity $\sigma_{zz}(E_z)\equiv
j_z/E_z$ could be best understood in the investigation of the
differential conductivity
\be \sigma_\mathrm{diff}(E_z)\equiv\frac{\mathrm d j_z}{\mathrm d
E_z}.
\ee
 In order to derive it we have to differentiate the general
formula for the current in an external magnetic field
Eq.~(\ref{j_final_h}) and use the well-known properties of the
modified Bessel functions
\be
2\frac{d I_{m}(z)}{d z}= I_{m+1}(z)+ I_{m-1}(z)
\ee
which can be found in any special functions manual, for example see
Ref.~\onlinecite{Korn-Korn:61} or Ref.~\onlinecite{Wolfram:99} and
references therein.
We turn to differentiation with respect to the dimensionless
electric field $\mathcal P$, taking into account the value of
the decay rate $\tau_0$ that follows from the BCS theory
\be
\frac{\partial \mathcal{P}}{\partial E_z}=\frac{es\tau_0}{\hbar}
,\qquad
\tau_0^\mathrm{(BCS)}=\frac{\pi\hbar}{16T_c}.
\ee
From Eq.~(\ref{j_final_h}) it is straight forward that the electric
field-dependent part in the differential conductivity is given by
the expression
\begin{eqnarray}\nn
&&\frac{\partial}{\partial \mathcal{P}}\left[\sin(\mathcal{P}\tilde u)
I_1\left(\frac{r\sin(\mathcal{P}\tilde u)}{2\mathcal{P}}\right)\right]\\
&&=\frac{r\tilde s}{4\mathcal{P}^2}
\left(\mathcal{P}\tilde u\tilde c-\tilde s\right)
\left(I_0+I_2\right)
+\tilde u \tilde cI_1,
\end{eqnarray}
where for the sake of brevity we have introduced the notations
\be \tilde s=\sin(\mathcal{P}\tilde u),\quad \tilde
c=\cos(\mathcal{P}\tilde u),\quad I_m=I_m\left(
\frac{r\sin(\mathcal{P}\tilde u)}{2\mathcal{P}}\right). \ee
Having in mind the considerations above it can be easily shown that
the general formula for the differential conductivity takes the final form
\begin{eqnarray}
\label{sigma_diff_h}\nn
\sigma_\mathrm{diff}(\bar\epsilon,\mathcal{P},h)
\!\!&=&\!\!\frac{e^2\xi_{c}^2(0)}{16\hbar s\xi_{ab}^2(0)}
\!\int_{\tilde u=0}^\infty \mathrm d\tilde u \mathrm
e^{-(\bar\epsilon + r/2)\tilde u} \frac{(1-\mathrm e^{-c\tilde
u})h}{\sinh h\tilde u}
\\
&&\times \!\!\left[\frac{r\tilde s}{4\mathcal{P}^2}
\left(\mathcal{P}\tilde u\tilde c
-\tilde s\right)\left(I_0+I_2\right)
+\tilde u \tilde cI_1\right]\!.
\end{eqnarray}
This is the equation we are going to use in order to analyze the
presence of negative differential conductivity for given samples.
Hereof we may derive the expressions for $\sigma_\mathrm{diff}$ in
the special cases of weak and strong magnetic fields.

In the absence of external magnetic field the limit $h\rightarrow 0$ gives
$h/\sinh h\rightarrow 1$ and we obtain
\begin{eqnarray}
\label{sigma_diff}\nn
\sigma_\mathrm{diff}(\bar\epsilon,\mathcal{P})
\!\!&=&\!\!\frac{e^2\xi_{c}^2(0)}{16\hbar s\xi_{ab}^2(0)}
\!\int_{\tilde u=0}^\infty
\mathrm d\tilde u \mathrm e^{-(\bar\epsilon + r/2)\tilde u}
\frac{1-\mathrm e^{-c\tilde u}}{\tilde u}
\\
&&\times \!\!\left[\frac{r\tilde s}{4\mathcal{P}^2}
\left(\mathcal{P}\tilde u\tilde c
-\tilde s\right)\left(I_0+I_2\right)
+\tilde u \tilde cI_1\right]\!,
\end{eqnarray}
which may be derived straight from Eq.~(\ref{j_final}).

For strong magnetic fields we take into account only the contribution of
the lowest Landau level and the differentiation of Eq.~(\ref{j_3D_B})
reads
\begin{eqnarray}
\label{sigma_diff_B_c2}
\sigma_\mathrm{diff}(\bar\epsilon_h,\mathcal{P})\!&=&\! \frac{\pi
e^2\xi_{c}^2(0)B_z}{4\hbar s\Phi_0} \int^\infty_0 \mathrm d \tilde u
\mathrm e^{-(\bar\epsilon_h+r/2)\tilde u}\\&& \times
\!\!\left[\frac{r\tilde s}{4\mathcal{P}^2} \left(\mathcal{P}\tilde
u\tilde c -\tilde s\right)\left(I_0+I_2\right) +\tilde u \tilde
cI_1\right]\!.\nn
\end{eqnarray}
If we want to return to the one-dimensional differential conductivity
for strong magnetic fields close to the phase curve we only have to
omit the number of magnetic threads $B_z/\Phi_0$ in the upper equation,
which gives
\begin{eqnarray}
\label{sigma_diff_1D}
\sigma_\mathrm{diff}^\mathrm{(1D)}(\bar\epsilon,\mathcal{P})\!\!\!&=&\!\!\!
\frac{\pi e^2\xi_{c}^2(0)}{4\hbar s}\frac{}{}
\!\!\!\int^\infty_0 \mathrm d\! \tilde u
\mathrm e^{-(\bar\epsilon+r/2)\tilde u}\\
&&\times \!\!\left[
\frac{r\tilde s}{4\mathcal{P}^2}
\left(\mathcal{P}\tilde u\tilde c
-\tilde s\right)\left(I_0+I_2\right)
+\tilde u \tilde cI_1\right]\!.\nn
\end{eqnarray}
%

%%%%%%%%%%%%%%%%%%%%%%%%%%%%%%%%%%%%%%%%%%%%%%%%%%%%%%%%%%%%%%%%%
Of course the calculated derivative $(\partial j_z/\partial
E_z)_{\bar \epsilon}$ does not describe the whole effect. We have to
take into account also $(\partial\bar \epsilon/\partial
E_z)_{\epsilon}$ and the increase of the sample's temperature above
the ambient temperature
\be \mathcal{R}_\mathrm{therm}(T-T_\mathrm{amb}) =
\mathcal{V}j_zE_z, \ee
due to the thermal resistance. The ``chemical'' reaction of pairing
of normal charge carriers $e+e\rightarrow e^*$ creates also a
decreasing of the number of normal charge carriers, and in the Drude
formula, for example, we have to take into
account\cite{Larkin-Varlamov:05} the density of state's corrections
(DOS)
$\sigma_\mathrm{norm}=
(n_e-2n(\bar\epsilon,\mathcal{P}))e^2\tau_\mathrm{norm}/m_e.$

Al those effects are however smaller compared to the dominating
Aslamazov-Larkin conductivity which qualitatively can describe the
appearance of new physical effects such as NDC, for example.
%%%%%%%%%%%%%%%%%%%%%%%%%%%%%%%%%%%%%%%%%%%%%%%%%%%%%%%%%%%%%%%%%%%%%

In order to better understand the nature of NDC in supercooling
regime in the next section we will analyze the Green functions of
the Boltzmann equation.

%%%%%%%%%%%%%%%%%%%%%%%%%%%%%%%%%%%%%%%%%%%%%%%%%%%%%%%%%
\section{Green functions of the Boltzmann equation}

Now let us paint some Cooper pairs in \emph{green} in order to trace
their motion i.e. in order to analyze the influence of the random
noise on the momentum distribution of the Cooper pairs we will
replace the constant income term in the Stoss-integral $n_0/\tau_0$
with a $\delta$-function inhomogeneous term in
Eqs.~(\ref{Boltzmann_2006}) and (\ref{Boltzmann_1872})
\be \frac{n_0}{\tau_0}\rightarrow \frac{2\pi\hbar}{L} \delta(t)
\delta (p-p_0). \ee
For simplicity we will consider 1D case.

Before the moment of creation there are no Cooper pairs
$n_p(t<0)=0.$ Then immediately after the influence we have
$\delta$-like initial distribution
\be
n_p(t=+0)=\delta_{p,p_0}
=\frac{2\pi\hbar}{L}
\delta (p-p_0).
\ee
This distribution is normalized to have one Cooper pair in the
initial moment
\be
\mathcal{N}(t)=\sum_p n_p(t)
=L\int \frac{\mathrm d p}{2\pi\hbar} n_p(t),
\quad
\mathcal{N}(+0)=1
\ee
born by thermal fluctuations; pictorially speaking ``by the sea foam.''
According to the separation of variables in the Boltzmann equation
Eq.~(\ref{Boltzmann_2006})
in every momentum point the evolution of the distribution function
is independent
\be
n_p(u=t/\tau_0)=n_p(0)
\exp\left\{-\int_0^u 2\nu(u_3)\mathrm \, d u_3\right\}.
\ee
The substitution here of the decay rate from Eq.~(\ref{decay_rate})
gives
\be
\label{droplet_size}
\frac{\mathcal{N}(u)}
{\mathcal{N}(0)}
=\exp\left\{-\left(\epsilon+\frac{r}{2}\right)u
+\frac{r\sin(\mathcal{P}u)}{2\mathcal{P}}
\cos(\theta-\mathcal{P}u)
\right\}.
\ee
It is instructive to consider the 3D limit case of $|\epsilon|\ll
r;$ formally one can consider the unrealistic for the cuprates case
of $r\gg1.$ For small angles
\be
\theta = \frac{2sk}{\sqrt{r}}\ll1,
\quad
\mathcal P=\frac{f}{\sqrt{r}},
\quad
\mathcal P u\ll1
\ee
the Taylor expansion of the trigonometric functions in
Eq.~(\ref{droplet_size}) gives
\be
\label{droplet_size_3D}
\frac{\mathcal{N}(u)} {\mathcal{N}(0)}
=\exp\left\{
-(\epsilon+q^2)u +qfu^2-\frac{1}{3}f^2u^3
\right\}.
\ee
The simplest example is to consider zero initial momentum $q=
\xi_c(0)p_0/\hbar=0.$ In the supercooling regime where $\epsilon<0$
in the beginning the number of Cooper pairs increases. It is a
typical lasing process -- increasing of the number of coherent Bose
particles in one mode; imagine a drop of rain increasing by the
condensation in a humid atmosphere. Finally, however, the
$\frac{1}{3}f^2u^3$ term dominates in the argument of the exponent
and $\mathcal N(\infty)=0;$ one can say that the droplet is
evaporated by the big kinetic energy like a meteorite falling in the
earth atmosphere. The droplets are not smeared during their lifetime
\be
n_p(t)=
\frac{2\pi\hbar}{L}
\frac{\mathcal{N}(u)}{\mathcal{N}(0)}
\delta (p-p_0).
\ee
They have just a drift in the kinetic momentum space. This drift
related to the electric current is analogous to the rain fall
created by the earth acceleration.
\be
n(\mathbf{p},t)=
\frac{2\pi\hbar}{\mathcal{V}}
\frac{\mathcal{N}(u)}{\mathcal{N}(0)}
\delta (\mathbf{p}-\mathbf{p}_0-e^*\mathbf{E}t).
\ee
In upper \emph{Green} function solution of the Boltzmann
equation Eq.~(\ref{Boltzmann_1872}) we have recovered the 3D variables.

Now we can understand qualitatively the mechanism of creation of NDC
for small electric fields in a supercooled superconductor. Let us
take for illustration $p_{0}=0$ case in Eq.~(\ref{droplet_size_3D}).
For $\epsilon<0,$ $h=0,$ and $|\epsilon|\ll r$ this function has a
maximum\cite{MPI:02}
\be
\frac{\mathcal{N}(u_\mathrm{max})}{\mathcal{N}(0)}
=\exp\left\{
\frac{2}{3}\frac{(-\epsilon)^{3/2}}{f}
\right\}
\ee
at
\be
u_\mathrm{max}=\frac{\sqrt{(-\epsilon)}}{f}.
\ee
The maximal increase of the ``weight'' of the Bose droplet
$\mathcal{N}(u_\mathrm{max})/\mathcal{N}(0)$ can reach big values
for small enough electric fields
\be
f\ll (-\epsilon)^{3/2},
\ee
or
\be
e^*E\tau(\epsilon)\xi(\epsilon)/\hbar\ll1,
\quad
\tau(\epsilon)=\frac{\tau_0}{|\epsilon|},
\quad
\xi(\epsilon)=\frac{\xi(0)}{|\epsilon|^{1/2}}.
\ee
Decreasing of the electric field increases the maximal size of the
Cooper pair droplets and the electric current which is proportional
to the number of particles in the droplets. This precursor of the
Bose condensation creates the NDC. Our preliminary numerical
calculations for YBa$_2$Cu$_3$O$_{7-\delta}$ (Y:123) taking the
parameters from Ref.~\onlinecite{Lang:05}: $s=1.17$~nm,
$\xi_{ab}(0)=1.2$~nm, $\xi_{c}(0)=0.14$~nm, $T_c=92.01 \;
\mathrm{K},$ $T_c^\mathrm{(ren)}=86.9$~K, $\kappa_\mathrm{GL}=70,$
$1/\sigma_{zz}^{\mathrm{(norm)}}=5\:\mathrm{m\Omega cm},$ $c=0.5$
demonstrated existence of NDC; this gives the hope that NDC can be
experimentally observed for out-of-plane transport in Y:123.

%%%%%%%%%%%%%%%%%%%%%%%%%%%%%%%%%%%%%%%%%%%%%%%%%%%%
\section{Analysis, discussion and conclusion}

Let us start with the formal discussion. In case of big enough
overcooling $\epsilon + r/2<0,$ when $\epsilon<0$ and $|\epsilon|>r$
the integral for the current Eq.~(\ref{j_final}) is divergent. This
means that the electric field cannot prevent the appearance of a
superconducting condensation, as is the case of volume
superconductors. The reason for this is that the one-dimensional
zone in the energy spectrum Eq.~(\ref{E_kin}) has a finite width
$a_0r,$ in contrast to the LD model, which shall be obtained if we
take the continual limit of $r\gg 1.$ However, if the LD parameter
is small enough we have to take into account the interaction between
the fluctuation modes, described by the nonlinear term in the GL
theory. These effects become important when the reduced temperature
$\epsilon$ is comparable to the Ginzburg number for layered
superconductors
\be \epsilon_{\mathrm {Gi}}(T_c) =
\frac{1}{4\pi\xi^2_{ab}(0)s\Delta C},
\ee
where (see Ref.~\onlinecite{Mishonov-Penev:00},~Sec.~3.3,~Eq.~169)
$\Delta C$ is the jump of the heat capacity, extrapolated from the
critical behaviour of the heat capacity $C(T)$ as a fitting
parameter, or can be evaluated by the electrodynamic properties
\be
T_c\Delta C=\frac{1}{8\pi^2\mu_0}
\left(\frac{\Phi_0}{\lambda_{ab}(0)\xi_{ab}(0)}\right)^2.
\ee
In the Maxwell-Hartree SCA the total volume density of the
fluctuation Cooper pairs leads to an effective heating (see
Ref.~\onlinecite{Mishonov-Penev:00},~Sec.~3.3,~Eqs.~(164) and
(173)). As a whole, inclusion of the SCA does not change the
qualitative behavior of the obtained results. In overcooled
superconductors $\epsilon <0,$ for example, we expect the appearance
of a negative differential conductivity~(NDC) with the decreasing of
the constant electric field. Whether this will lead to formation of
a domain structure of normal and superconducting layers or to
existence of electric oscillations is a question of further
analysis. Our theory, however, predicts that this phase transition
when $T<T_c$ will take place via annulation of the differential
conductivity. In this respect the layered superconductors give an
advantage as, because of the strong anisotropy, the normal
conductivity perpendicular to the layers is too small. Thus the
sample will not heat intensively, which gives the opportunity to be
observed the nonlinear effects of a strong electric field over the
fluctuation conductivity. These nonlinear effects can be observed as
an amplification of the current harmonics generation as we reach the
critical region. Systematical investigation of these harmonics will
lead to specification of the parameters in the TDGL theory and to
clarification of the kinetics of the superconducting order
parameter. Here we would like to consider the possible application
of our theory for different layered cuprates. For instance, for the
extremely anisotropic Bi$_2$Sr$_2$CaCu$_2$O$_{8+x}$ (Bi:2212) the
value of the LD parameter $r$ is exceptionally small. The heating
from the electric field applied in the ``dielectric'' $z$-direction
is very weak, but the area where current harmonics can be observed
is very narrow as well. Because of this their observation would
require samples of extraordinary quantity. Wider and easier for
observation would be the nonlinear region for the cuprates with a
moderate anisotropy like Y:123. For most favorable we consider the
cuprate Ta$_2$Sr$_2$CuO$_6$ (Ta:2201) where a coherent Fermi surface
is observed. When the anisotropy is small and $r$ parameter bigger,
the half-width of the $z$-zone $\frac{1}{2}a_0r=\hbar^2/m_c^2s^2$ is
bigger and then the electric field $E_z$ causes more significant
increase of the kinetic energy of the Cooper pairs
$\frac{1}{2}a_0r(1-\cos p_{z}s/\hbar).$ Because of this, for
instance, the overcooling region when $\epsilon<0$ and
$\mathcal{P}\neq0$ would be wider and more easily observed. Exactly,
in such more easily reached by the experiment conditions, we hope to
observe the annulation of the total differential conductivity. For
this purpose submicron high qualitative films of Ta:2201 are
necessary to be used. The investigation of the harmonics generation
above $T_c$ is a useful beginning for the systematical research of
the nonlinear effects of the strong electric field over the
fluctuation current in $z$-direction.

The experiment which we advocate is principally very simple, but
according to the best of our knowledge not done yet. The fluctuation
conductivity have to be investigated in a \emph{voltage biased
circuit}; a constant (DC) voltage has to be applied. A significantly
smaller AC voltage has also to be applied to the sample. The DC
current gives the conductivity and a measurement with a smaller AC
current gives the differential conductivity. The circuit should be
similar to the devices invented for investigation of the NDC in
tunnel diodes. The DC voltage has to be added to the normal phase
above $T_c$ and the simplest experiment which we suggest is to
measure the temperature dependence of the AC current (differential
conductivity) as a function of the temperature. The theory reliably
predicts annulation of the differential conductivity at cooling
below $T_c.$ What exactly will happen at further cooling is
difficult to be predicted experimentally: it could be NDC if the
space homogeneity is conserved, creation of layered domain structure
of normal and superconducting layers or complicated dynamic
phenomena. Different instabilities can be triggered by small
perturbations including the nature of contacts and chemical
treatment of the surfaces. One thing is sure: indispensably, new
physics will happen in the area below the annulation of the
differential conductivity. The realization of the simplest possible
scenario of creation of NDC will open the technological perspective
to create a new type of high-frequency oscillators operating in the
THz gap.\cite{Mishonov-Mishonov:05} While in
Ref.~\onlinecite{Mishonov-Mishonov:05} is analyzed the case of bulk
superconductors in the present work we demonstrated that for
electric currents applied in the ``dielectric'' $c$-direction the
dissipated power could be orders of magnitude smaller and perhaps
Josephson coupled superconductors can realize the first technical
applications of NDC in superconductors; it was one of the
motivations to perform the present work.

Few words we wish to add concerning the theoretical notions used in
the present work. It seems very strange that Boltzmann kinetic
approach has still limited usage in the physics of paraconductivity
as it was for the normal conductivity hundred years ago. Just Green
functions of the Boltzmann equation give the transparent picture
what is happening in homogeneous electric fields. Separation of the
variables in optical gauge with zero electric potential $\varphi =0$
reduces the kinetic equation to an ordinary differential equation.
Returning to the kinetic momentum space gives the drift of the
fluctuation Cooper pairs created by the electric field.

Let us continue the discussion of physics in supercooling regime.
Near to the annulation point of the differential conductivity the
paraconductivity is not a small excess perturbation but will become
comparable with the fluctuation conductivity. In the suggested
experiment of applied DC electric field and small AC voltage one can
observe in principle the AC component of the fluctuation
magnetization $M_z.$ This physical effect is described by
magnetoelectric susceptibility $\chi_\mathrm{ME}=\partial
M_z/\partial E_z;$ this theory will be the subject of a future
research. Measurements of magnetization and paraconductivity in one
and the same sample can lead to exact determination  of the time
constant $\tau_0.$ This is the most important parameter of TDGL
theory which reveals in part the nature of superconductivity and
pairing mechanism. The $\pi/8$ multiplier in Eq.~(\ref{time_const})
is only weak coupling BCS result, but as we already mentioned, for
references of relevant works see Sec 4.2 of the
review,\cite{Mishonov-Penev:00} the experimental accuracy now is not
enough to observe reliably any deviations from this BCS $\pi/8$
value. There is no consensus that BCS theory is directly applicable
for high-$T_c$ cuprates and one might expect some correction
multiplier $\tau_\mathrm{rel}$ to the life time of the order of one.
The relative life time can give important information for the
pairing mechanisms in cuprates.

From qualitatively point the numerical value of $\tau_0$ is not so
important because only fixes the scale for the graphical
presentation of the experimental data. In the chosen dimensionless
units the current voltage curves should be universal and the most
important property of the current voltage curves is the predicted
annulation of the differential conductivity at supercooling below
$T_c.$ Demonstration of this annulation and possible NDC in
out-of-plane transport can trigger significant technical
applications\cite{Mishonov-Mishonov:05} and we hope that
experimental search of NDC in superconductors can start in the
nearest future.

\acknowledgments The scientific discussions, support, critical
reading of the manuscript, and correspondence with A.~Varlamov,
D.~Damianov, V.~Mishonova and W.~Lang are highly appreciated.

%%%%%%%%%%%%%%%%%%%%%%%%%%%%%%%%%%%%%%%%%%%%%%%%%%%%%%%%%%%%%%%%%%%%%
\bibliographystyle{asprev}

\end{document}